\documentclass[12pt]{elsart}

\usepackage{graphicx}
\usepackage{subfigure}
\usepackage{amsmath}
\usepackage{amssymb}
\usepackage{cite}

\journal{Physics Letters A}

\begin{document}

\begin{frontmatter}

\title{Critical properties of the frustrated Ising model on a honeycomb lattice: A Monte Carlo study}
\author{M. \v{Z}ukovi\v{c}\corauthref{cor}}
\ead{milan.zukovic@upjs.sk}
\address{Institute of Physics, Faculty of Science, P.J. \v{S}af\'arik University\\ Park Angelinum 9, 041 54 Ko\v{s}ice, Slovakia}
\corauth[cor]{Corresponding author.}

\begin{abstract}
Critical and in the highly frustrated regime also dynamical properties of the $J_1-J_2$ Ising model with competing nearest-neighbor $J_1$ and second-nearest-neighbor $J_2$ interactions on a honeycomb lattice are investigated by standard Monte Carlo and parallel tempering simulations. The phase boundary is determined as a function of the coupling ratio for the phase transition between the paramagnetic and ferromagnetic states within $R \equiv J_2/|J_1| \in [-1/4,0]$. It is confirmed that at least for $R \geq -0.2$ the transition remains second-order and complies with the standard Ising universality class. In the highly frustrated regime of $R < -0.2$ and low temperatures the system tends to freeze to metastable domain states, separated by large energy barriers, which show extremely sluggish dynamics. The resulting huge equilibration and autocorrelation times hinder the analysis of critical properties by using standard Monte Carlo techniques. The parallel tempering approach facilitates much better tunneling through the energy barriers, nevertheless, the convergence to the equilibrium remains beyond the reach even for relatively small system sizes and thus the character of the transition in this region remains to be determined. 
\end{abstract}

\begin{keyword}
Ising model \sep honeycomb lattice \sep frustration  \sep phase transition

\end{keyword}

\end{frontmatter}

\section{Introduction} 
Besides Ising antiferromagnetic models on non-bipartite, such as triangular, lattices, Ising models on bipartite lattices with competing further-neighbor antiferromagnetic interactions are perhaps the simplest examples of frustrated spin systems. In the latter case, the degree of frustration can be adjusted by tuning the ratio between the nearest- and further-nearest interactions and can lead to various interesting phenomena. In particular, the two-dimensional Ising antiferromagnet on a square lattice has a long history of investigations~\cite{nigh77,krin79,land80,bind80,land85,oitm81,blot87,gryn92,mora93,mora94,lope99,mala06,anjo08,kalz08,kalz11,jin12,jin13,boba15}. It has been found that competition between the (ferromagnetic or antiferromagnetic) nearest-neighbor interaction $J_1$ and the antiferromagnetic second-nearest-neighbor interaction $J_2$ leads to gradual reduction of the critical temperature until there is no long-range ordering (LRO) at any finite temperature at $R \equiv J_2/|J_1| =-1/2$~\cite{land80,gryn92,mora93,mora94}. Most of the earlier studies concluded that the transition remains second-order with the critical exponents belonging to the Ising universality class, however, the more recent cluster mean-field calculation~\cite{jin13} and effective-field theory (EFT) with correlations~\cite{boba15} suggested the possibility of the first-order transition near $R =-1/2$. For $R <-1/2$ the system shows a phase transition to a peculiar striped state. The character of the phase transition in this regime of a relatively strong $J_2$ has been even more controversial. A scenario proposed by a series of earlier studies predicted a second-order transition with non-universal critical exponents for any $R < -1/2$ (see Ref.~\cite{mala06} and references within), while some more recent approaches favored a first-order transition for $R^* < R < -1/2$ and a continuous one only for $R < R^*$~\cite{mora93,mora94,anjo08,kalz08,kalz11,jin12,boba15}. Nevertheless, even the value of $R^*$ in the latter cases has been a subject of controversy with rather different estimates ranging from the early $-1.1$~\cite{mora93} down to the latest $-0.67$~\cite{jin12}. The effects of a transverse field~\cite{boba18,sadr18,kell19} and stacking of the 2D layers in a vertical direction~\cite{anjo07,godo20} on the critical properties the model have also been investigated.\\
\hspace*{5mm} Considering the amount of work done on the frustrated $J_1-J_2$ model on the square lattice, it is surprising that the corresponding model on the honeycomb lattice has received much less attention~\cite{hout50,kudo76,kats86,boba16,zuko20,cort20,acev21}. In fact, due to the smaller coordination number the latter frustrated model might display even more interesting effects, similar to its Heisenberg counterpart~\cite{cabr11,zhan13}. The critical behavior of this model on the honeycomb lattice was studied by the EFT~\cite{boba16} and very recently also by interesting machine learning approaches~\cite{cort20,acev21}. Nevertheless, the order of the transition was only investigated in the former approach with not quite conclusive findings. In particular, within the interval $-1/4 < R < 0$ a single-spin cluster EFT approximation produced only the second-order phase transition but a tricritical behavior with a crossover to the first-order behavior at sufficiently large $J_2$ ($R \lesssim -0.1$) was obtained when larger spin cluster sizes were used. We note that the EFT also predicted the first-order phase transitions within $-1/2 < R < 0$ for the square lattice, however, such behavior was not reproduced by more sophisticated, such as Monte Carlo, approaches. \\
\hspace*{5mm} Therefore, to clarify the critical behavior of the frustrated $J_1-J_2$ model on the honeycomb lattice more studies applying other techniques are desirable. In the present letter we approached the model by using both standard Monte Carlo and parallel tempering simulations. Our findings do not support the presence of  the first-order transition, obtained by the EFT for $R \lesssim -0.1$, but rather suggest that the transition stays second-order down to at least $R \approx -0.2$. Nevertheless, a crossover to the first-order behavior for $R<-0.2$ cannot be completely ruled out.

\section{Model and Methods}
The $J_1-J_2$ Ising model on the honeycomb-lattice with nearest-neighbor, $J_1$, and second-nearest-neighbor, $J_2$, exchange interactions can be described by the Hamiltonian 
\begin{equation}
\label{hamiltonan}
\mathcal{H} = -J_1\sum_{\langle i,j\rangle}s_i s_j - J_2 \sum_{\langle i,k\rangle}s_i s_{k},
\end{equation}
where $s_i = \pm 1$ is the Ising spin variable at the $i$th site and the summations $\langle i,j\rangle$ and $\langle i,k\rangle$ run over all nearest and second-nearest spin pairs, respectively. Frustration is induced by the competing ferromagnetic ($J_1>0$) and antiferromagnetic ($J_2<0$) couplings. In the following, we fix the energy scale by putting $J_1 = 1$ and absorb the Boltzmann constant in temperature $T$, which then will be measured in units of $J_1$.\\
\hspace*{5mm} The model is first approached by standard Monte Carlo (MC) simulations, using the Metropolis algorithm. The system sizes range from $L=24$ up to $144$, with periodic boundary conditions. For thermal averaging of the calculated thermodynamic quantities at each value of the reduced temperature $T$ we consider in standard runs  $2 \times 10^5$ and in long runs $10^7$ MC sweeps (MCS). Typically, before we start collecting the time series for the thermal averaging we discard the initial twenty percent of those numbers in order to bring the system to thermal equilibrium. To obtain temperature dependencies of the calculated quantities, the simulations start from high temperatures from random initial states and then the temperature is gradually decreased. The next simulation starts from the final configuration obtained at the previous temperature to ensure that the system is maintained close to the equilibrium in the entire temperature range. In the highly frustrated region of $R \lesssim -0.2$, when  MC simulations tend to freeze in metastable domain states, we also perform runs by starting at low temperatures from ferromagnetic initial states, in which case the temperature is gradually increased.\\
\hspace*{5mm} With the increasing frustration ($R \to -1/4$ and $T \to 0$) the system tends to freeze in metastable domain states and the standard MC simulations turn out to be either very inefficient or completely failing in reaching the stable thermal equilibrium state. In effort to alleviate these difficulties we applied the parallel tempering (PT) or replica exchange method~\cite{huku96}, which more efficiently overcomes energy barriers by a random walk in temperature space and allows exploration of complex energy landscapes. In such a case, we run simulations at $N_T$ temperatures (replicas) in parallel and propose $10^5$ swaps between replicas followed by $100$ MCS over the complete lattice. The swap acceptance rate between neighboring replicas is $ P(\beta_{i} \leftrightarrow \beta_{i+1})=\min\{1,\exp(\Delta \beta \Delta E)\}$, where $\Delta \beta =\beta_{i+1}-\beta_{i}$ and $\Delta E = E_{i+1} -E_{i}$ are differences between the neighboring inverse temperatures, $\beta_i=1/T_{i}$, and the energies of the corresponding configurations, respectively. The temperatures $T_i$ are chosen distributed symmetrically around the estimated transition temperature with their density logarithmically increasing away from the center. Such a setup focuses more on the simulation bottleneck in the vicinity of the transition temperature and ensures sufficient acceptance rates within the whole temperature interval. For estimation of statistical errors, we use the $\Gamma$-method~\cite{wolf04}.\\
\hspace*{5mm} The quantities of interest include the internal energy per spin $e \equiv E/N=\langle \mathcal{H} \rangle/N$, where $N=L^2$ is the number of spins, the fluctuations of which are related to the specific heat based on the formula
\begin{equation}
\label{spec_heat}
c =\frac{\langle \mathcal{H}^{2} \rangle - \langle \mathcal{H} \rangle^{2}}{NT^{2}},
\end{equation}
where $\langle \cdots \rangle$ denotes the thermal average. The peaks in the latter are useful for estimation of the phase transition temperatures. From the fluctuations of the magnetization $m \equiv \langle M \rangle/N=\langle \sum_{i=1}^{N}s_i\rangle/N$ one can obtain the susceptibility per spin $\chi$, using the formula
\begin{equation}
\label{eq.chi}\chi = \frac{\langle M^{2} \rangle - \langle M \rangle^{2}}{NT}.
\end{equation}
Furthermore, the following quantities are useful in finite-size scaling (FSS) analysis: the derivative and logarithmic derivative of $\langle M \rangle$ with respect to the inverse temperature $\beta=1/T$
\begin{equation}
\label{eq.D0}dm = \frac{\partial}{\partial \beta}\langle M \rangle = \langle M \mathcal{H} \rangle- \langle M \rangle \langle \mathcal{H} \rangle,
\end{equation}
\begin{equation}
\label{eq.D1}dlm = \frac{\partial}{\partial \beta}\ln\langle M \rangle = \frac{\langle M \mathcal{H} \rangle}{\langle M \rangle}- \langle \mathcal{H} \rangle.
\end{equation}
In the FSS analysis we employ the following scaling relations:
\begin{equation}
\label{eq.scal_chi}\chi_{max}(L) \propto L^{\gamma/\nu},
\end{equation}
\begin{equation}
\label{eq.scal_dlm}dlm_{max}(L) \propto L^{1/\nu},
\end{equation}
\begin{equation}
\label{eq.scal_m}m_{max}(L) \propto L^{-\beta/\nu},
\end{equation}
\begin{equation}
\label{eq.scal_dm}dm_{max}(L) \propto L^{(1-\beta)/\nu},
\end{equation}
\begin{equation}
\label{eq.scal_c1}c_{max}(L) = c_0+c_1\ln(L),
\end{equation}
\begin{equation}
\label{eq.scal_Tc}\beta_{max}(L) = \beta_c + a L^{-1/\nu},
\end{equation}
where $\beta_c$ is the inverse transition temperature and $\beta_{max}(L)$ are the inverse pseudo-transition temperatures, estimated as positions of the maxima of the above functions for a given $L$. Note that Eq.~(\ref{eq.scal_c1}) represents the logarithmic divergence of $c_{max}$, expected for the Ising universality class (critical exponent $\alpha=0$).

\section{Results} 
\begin{figure}[t!]   
\centering 
\subfigure{\includegraphics[scale=0.5,clip]{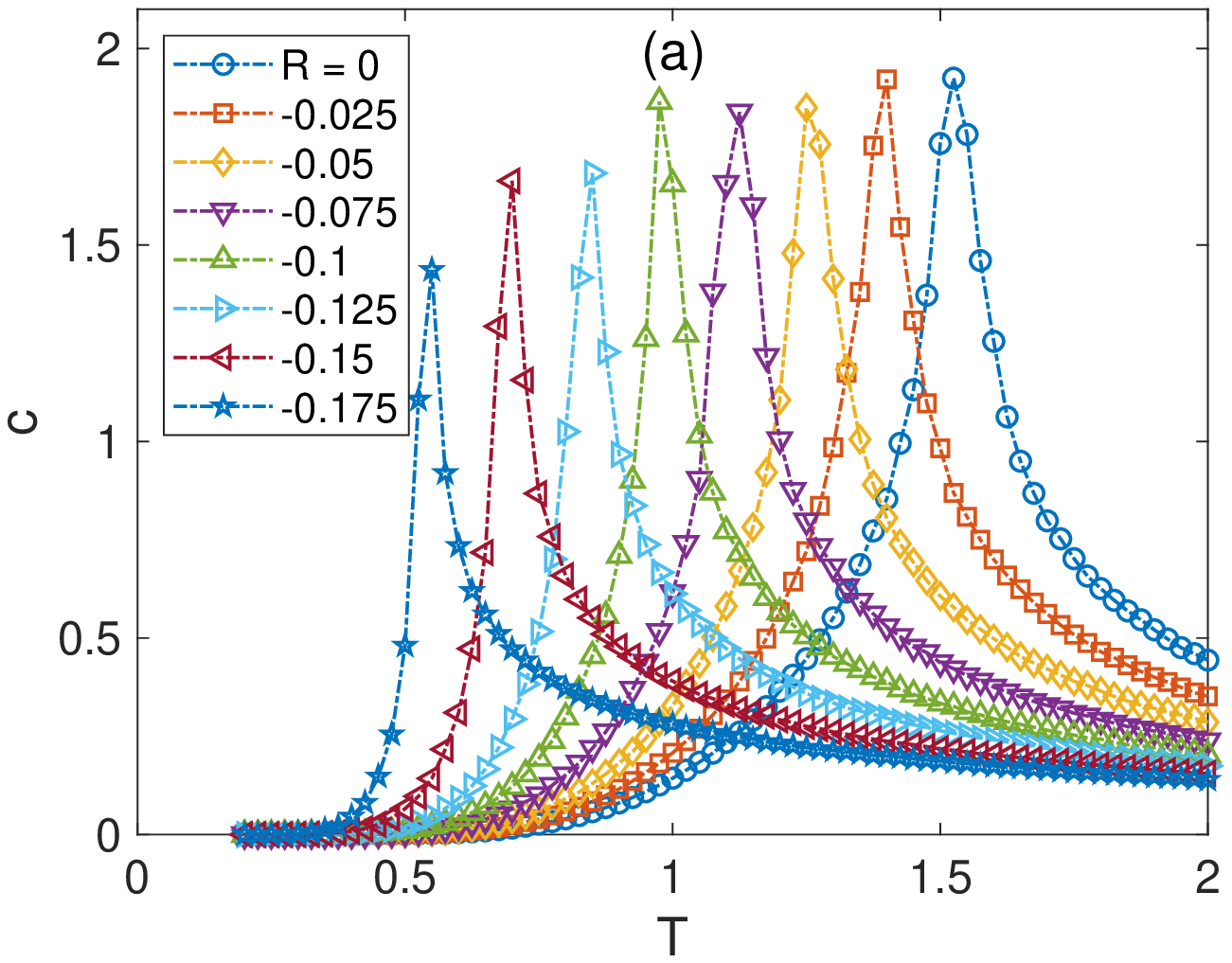}\label{fig:c-T_L48}}  
\subfigure{\includegraphics[scale=0.5,clip]{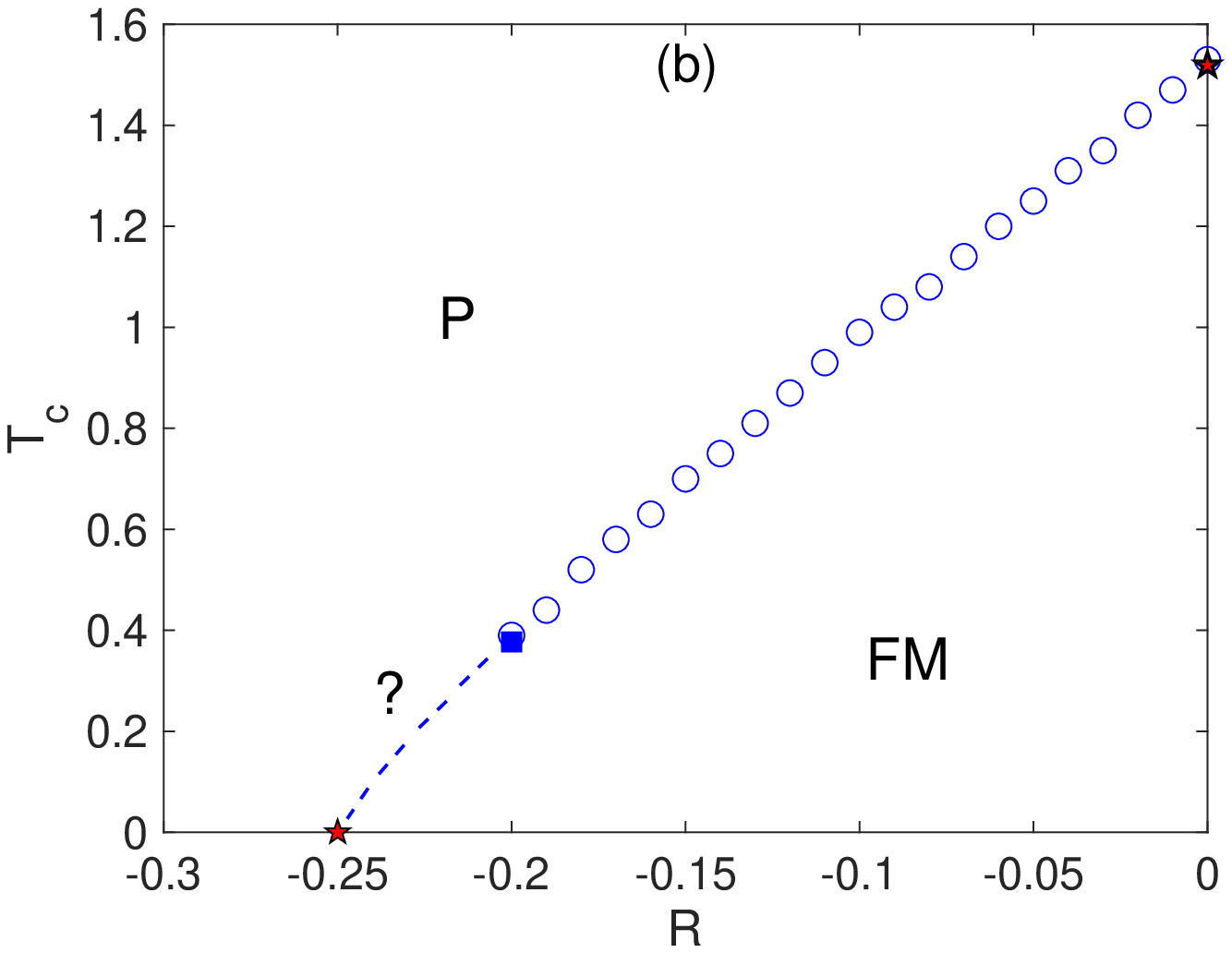}\label{fig:Tc-alp}}
\caption{(a) Specific heat temperature dependencies for various $R$ and $L=48$. (b) Phase diagram as a function of the exchange interaction ratio $R$. The (pseudo)transition temperatures, marked by circles, are estimated from the specific heat peaks, while the one at $R = -0.2$, marked by the filled square, is obtained from the FSS analysis. The limiting values at $[T_c,R] = [0,-1/4]$ and $[2/\log (2 + \sqrt{3}),0]$, marked by the filled stars, represent exact values. The dashed line is just a guide to the eye for the low-temperature phase boundary, which could not be reliably determined (see text).} 
\label{fig:PD}
\end{figure}

\begin{figure}[t!]   
\centering 
\includegraphics[scale=0.7,clip]{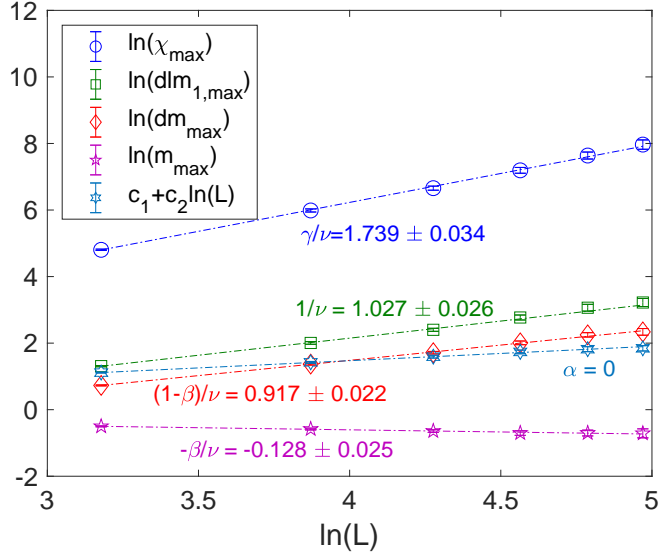}
\caption{Critical exponent ratios obtained in finite-size scaling analysis, from the scaling relations.} 
\label{fig:fss}
\end{figure}

\begin{figure}[t!]   
\centering 
\subfigure{\includegraphics[scale=0.5,clip]{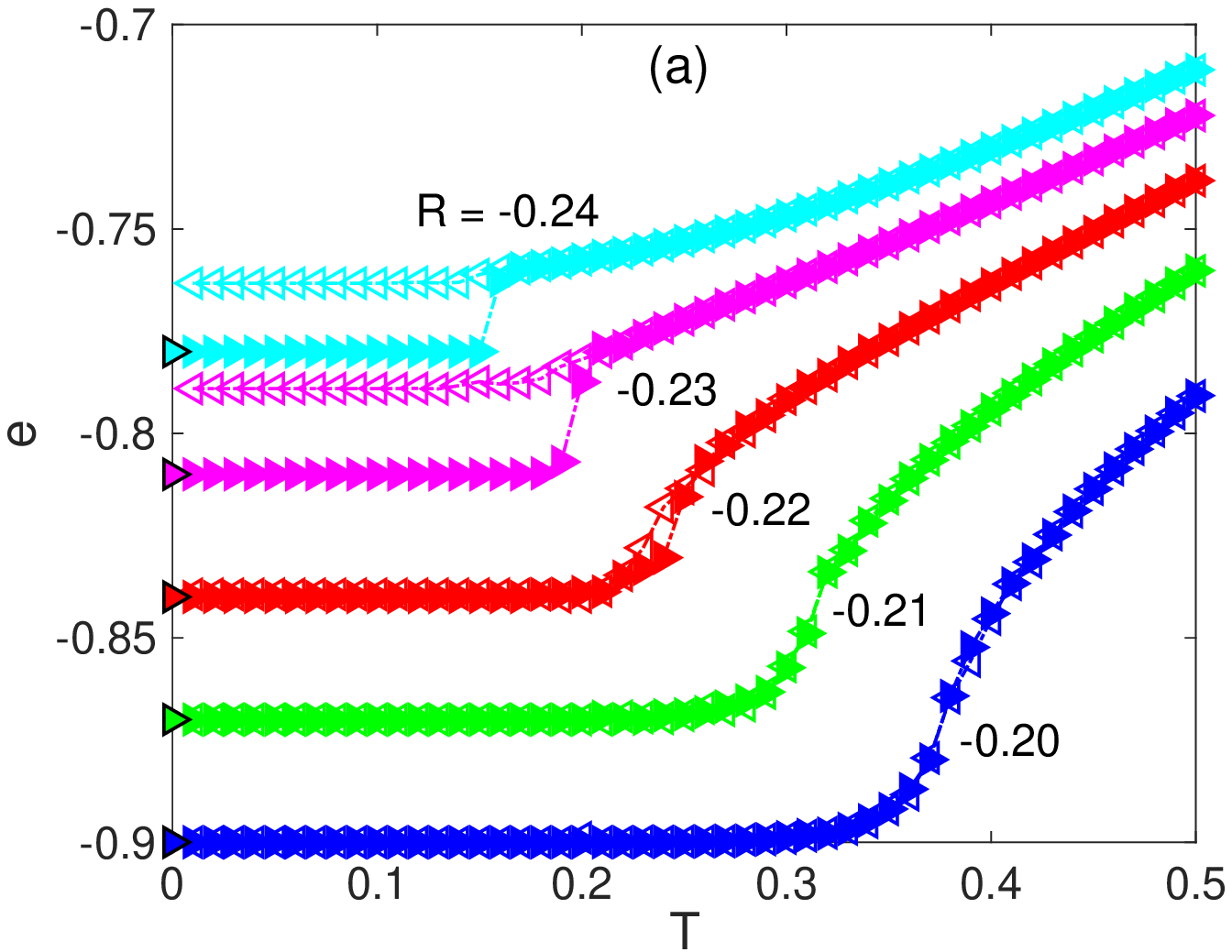}\label{fig:e-T_L48}}  
\subfigure{\includegraphics[scale=0.5,clip]{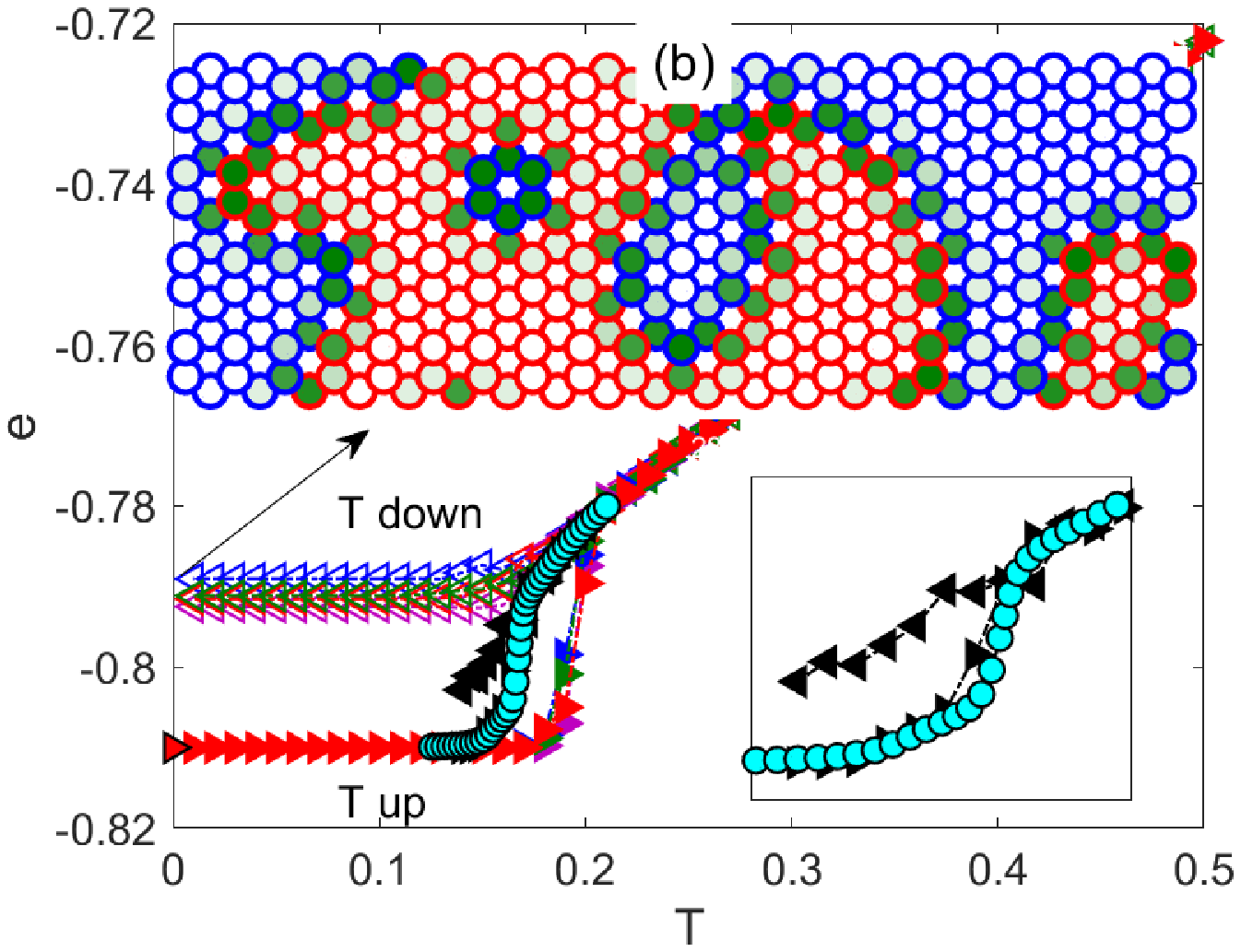}\label{fig:e-T_L48_alp-0_23}}
\caption{Temperature dependencies of the internal energy in the temperature decreasing (left-pointing triangles) and increasing (right-pointing triangles) processes, for (a) several values of $R \leq -0.2$ and (b) for $R =-0.23$ and several different MC runs. The filled black symbols represent the results from long MC runs, using $10^7$ MCS, and the black-framed cyan circles correspond to the PT results. The black-framed right-pointing triangles at $T=0$ represent the exact ground-state values. The inset  in (b) in the lower right corner zooms on the long MC and PT results and the upper inset shows the low-temperature spin snapshot of a metastable state with distinct spin-up (red circles) and spin-down (blue circles) domains. The shading of the spin circle interiors corresponds to deviation of the local energy of each spin from the ground-state value, with the white color representing zero.} 
\label{fig:e-T}
\end{figure}

\begin{figure}[t!]   
\centering 
\subfigure{\includegraphics[scale=0.5,clip]{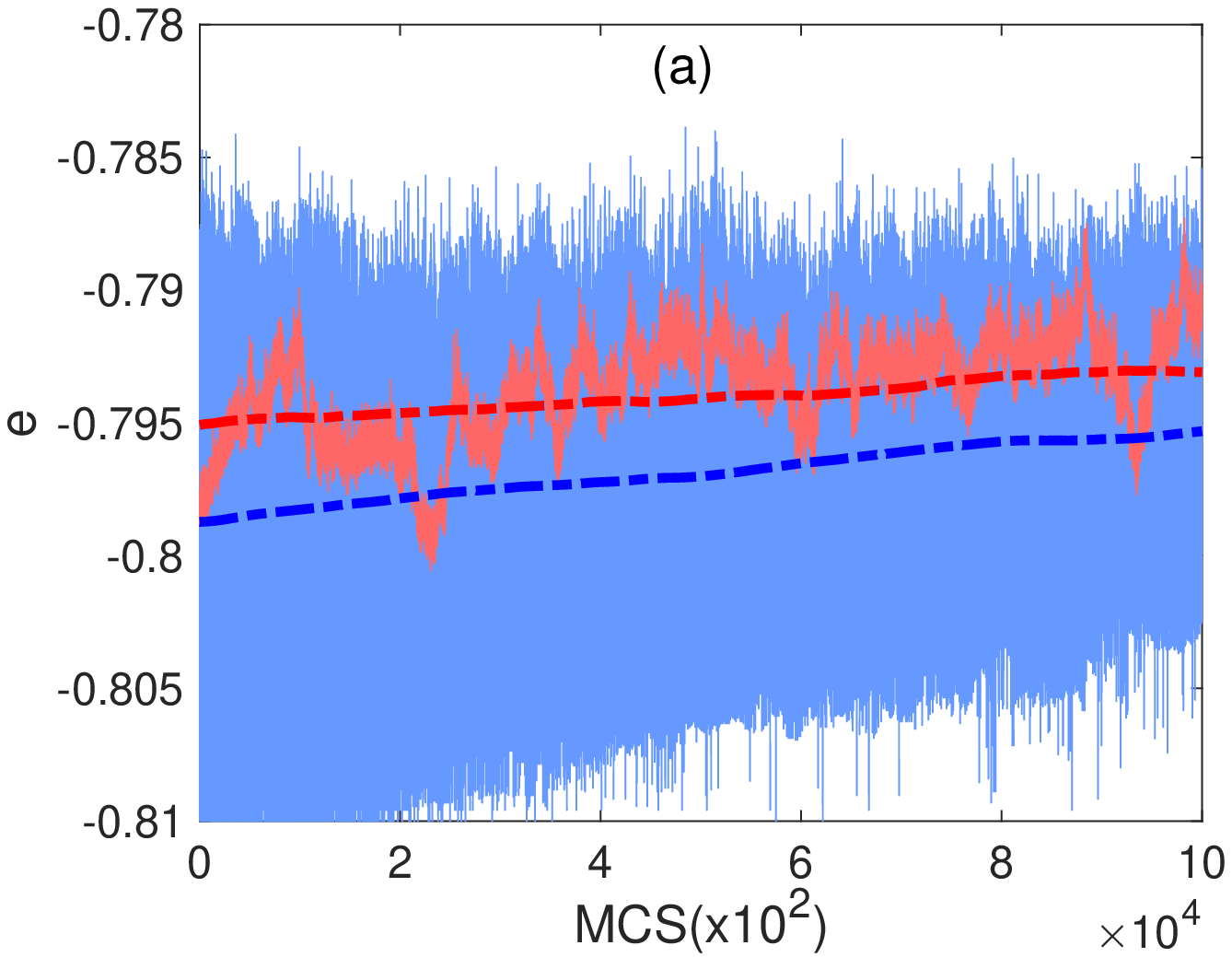}\label{fig:e-mcs_L48}}  
\subfigure{\includegraphics[scale=0.5,clip]{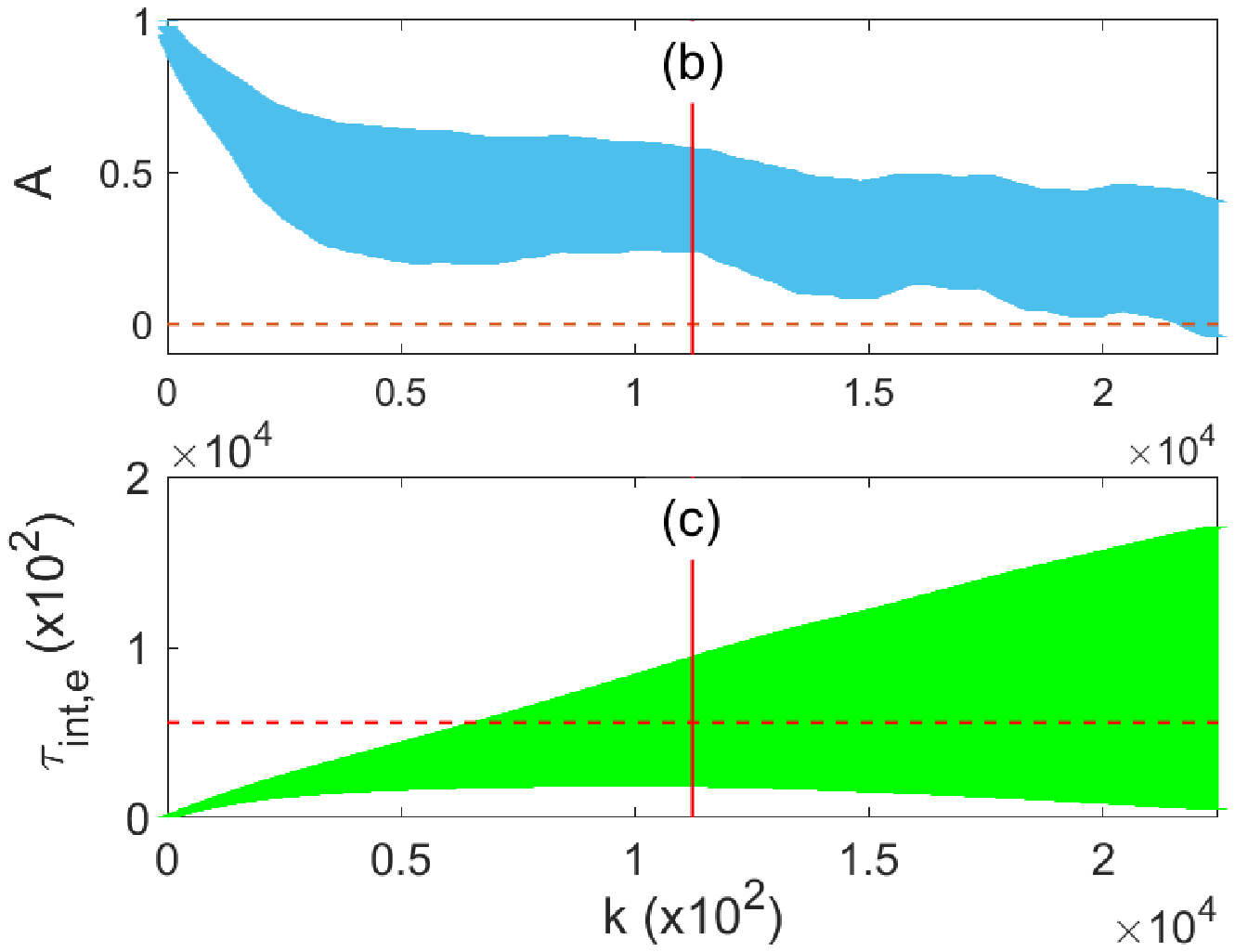}\label{fig:tau_int_e-L48_Tc}}
\caption{(a) Evolution of the energy near the transition point from the standard MC (thin red curve) and PT (thin blue curve) simulations. The corresponding thick dashed curves represent moving averages to highlight the trends in the respective time series. (b) Normalized autocorrelation function, $A(k)$, and (c) the integrated autocorrelation time of the energy, $\tau_{int,e}$  from the standard MC. All the results correspond to the simulations at $T = 0.168$, $L = 48$ and $R=-0.23$.} 
\label{fig:auto_corr}
\end{figure}

In the following we focus on determination of the phase diagram and the character of the phase transitions in the $J_1-J_2$ model for $R \in [-1/4,0]$. The (pseudo)transition temperatures can be estimated based on the peaks positions in temperature dependencies of various response functions, such the specific heat shown for different $R$ in Fig.~\ref{fig:c-T_L48}. As expected, the decreasing $R$  increases frustration, which is reflected in the transition temperatures being shifted to lower values. The resulting phase diagram in the $R-T$ plane is plotted in Fig.~\ref{fig:Tc-alp}. The circles represent the values estimated from the specific heat peaks and the values at the boundaries, i.e., $[T_c,R] = [0,-1/4]$ and $[2/\log (2 + \sqrt{3}),0]$, marked by the filled stars, represent the exact values in the ground state~\cite{kudo76,kats86} and in the absence of further-neighbor interactions~\cite{hout50}, respectively. In order to asses how the pseudo-transition temperatures, obtained for a selected finite lattice size, deviate from the thermodynamic limit values, at $R = -0.2$ we also include the critical temperature (filled square) obtained from the FSS analysis (Eq.~(\ref{eq.scal_Tc})). As one can see, both the value determined from FSS at $R = -0.2$ as well as the exact one at $R = 0$ are only slightly overestimated by the pseudo-transition temperatures determined from the specific heat peaks positions. \\
\hspace*{5mm} Another important information that can be extracted from the FSS analysis is related to the character of the phase transitions. In particular, there are examples of spin systems in which competing further-neighbor interactions can lead to the change of the universality class or the order of the transition. In Fig.~\ref{fig:fss} we present results of the FSS analysis performed at $R = -0.2$. The obtained critical exponents ratios, $1/\nu=1.027 \pm 0.026, -\beta/\nu=-0.128 \pm 0.025,(1-\beta)/\nu=0.917 \pm 0.022$, and $\gamma/\nu=1.739 \pm 0.034$, as well as the logarithmic divergence of the specific heat, described by the FSS relations (\ref{eq.scal_chi})-(\ref{eq.scal_c1}), indicate a fairly good correspondence to the Ising universality class. Nevertheless, the degree of frustration at $R = -0.2$ is relatively high, which translates to deterioration of the linear fits and consequently somewhat increased error bars. Eventually, for $-1/4 < R < -0.2$ a reliable determination of the transition character and the transition temperatures becomes impossible.\\
\hspace*{5mm} In Fig.~\ref{fig:e-T_L48} we present temperature dependencies of the internal energy in this problematic region for $L=48$. For each value of $R$ there are two curves, corresponding to the temperature-decreasing (left-pointing triangles) process starting at high temperatures from the paramagnetic state and the temperature-increasing (right-pointing triangles) processes starting at low temperatures from the ferromagnetic state. One can notice that for $R=-0.22$ the pair of the curves splits in the vicinity of the expected transition temperature (anomalous decrease) and for $R<-0.22$ the temperature-decreasing curves do not join the temperature-increasing ones at low temperatures but rather follow different trajectories. In such a case, the energy corresponding of the temperature-decreasing branch is always higher and never reaches the expected ground-state value $e_{GS}=-3(2R+1)/2$~\cite{boba16}, denoted by the black-framed right-pointing triangles at $T=0$. Therefore, apparently, in the temperature-decreasing processes the system gets trapped in some rather stable, nevertheless, only metastable states. It is worth noticing that below the temperature corresponding to the branch separation point the energy is practically constant with minimum fluctuations, i.e., the system practically freezes in some state.\\
\hspace*{5mm} By running multiple simulations, the energies of which are shown  in Fig.~\ref{fig:e-T_L48_alp-0_23} for $R=-0.23$, we confirmed that the temperature-decreasing processes can freeze in different metastable states (see the left-pointing triangle curves) but, particularly for lower values of $R$, they are very unlikely to reach the stable ferromagnetic state. Instead, they freeze in domain states with relatively large ferromagnetically ordered clusters of spins, separated by domain walls with the increased energy, as shown by in the upper snapshot of Fig.~\ref{fig:e-T_L48_alp-0_23}. The circles filled with darker colors represent spins at the domain walls and the color intensity shows how much they locally increase the internal energy with respect to the ground-state value (white circles). The relative stability of the domain states can be attributed to the sluggish domain wall dynamics, when approached by the standard MC simulation. When we increase the simulation time, we see some evolution towards the stable state but very slow one. The black triangles correspond to the runs with the simulation times 50-times longer, i.e., using $10^7$ MCS. Near the transition point the left- and right-pointing triangle curves are much closer to each other but, apparently, at least in the temperature-decreasing process the stable state still has not been reached (see also Fig.~\ref{fig:e-mcs_L48}). The cyan circles represent the energy values obtained from the PT simulations. They almost coincide with the values obtained in the long MC runs in the temperature-increasing process, albeit, near the transition some differences persist.\\ 
\hspace*{5mm} In Fig.~\ref{fig:auto_corr} we assess the thermalization dynamics near the transition point in both approaches. In Fig.~\ref{fig:e-mcs_L48} we present the evolution of the internal energy from the standard MC (red curve) and PT (blue curve) simulations. The highly-fluctuating thin curves represent the raw time series, while the smoother thick dashed curves represent moving averages. The latter serve to better detect possible trends, which indicate non-equilibrium regime. It is obvious that the PT considerably improved the sluggish dynamics, observed in the standard MC simulations. Owing to the logarithmically distributed temperatures, we were able to achieve reasonably high and almost flat replica swap rates at neighboring temperatures in the whole temperature interval, thus maximizing the number of round-trips between the two extremal temperatures and substantially improving equilibration of the system (notice that during the entire simulations $e_{PT}<e_{MC}$). Nevertheless, the persisting trends in both curves indicate that the equilibrium was not reached in either case. The left panel of Fig.~\ref{fig:auto_corr} demonstrates a gigantic autocorrelation time accompanied with huge statistical errors in its determination from the internal energy time series obtained from the standard MC simulations near the transition temperature. In particular, the upper panel (b) shows the normalized autocorrelation function, $A(k)$, and the lower panel (c) the running integrated autocorrelation time, $\tau_{int,e}(k)$. From the latter, it can be estimated that $\tau_{int,e}$ is more than $5 \times 10^5$ MCS even for as small lattice size as $L=48$. $\tau_{int,e}$ is expected to further increase with $L$ as power law, which hampers any reliable equilibrium simulations to be used in FSS analysis within realistic computational times.\\

\section{Conclusions}
Finally, using the standard Metropolis as well as PT simulations we have studied critical and dynamical properties of the $J_1-J_2$ Ising model with competing nearest-neighbor ferromagnetic $J_1$ and second-nearest-neighbor antiferromagnetic $J_2$ interactions on the honeycomb lattice. We attempted to determine the phase diagram as a function of the coupling ratio $R$, for $R \in [-1/4,0]$, where the phase transition from the paramagnetic to the ferromagnetic state is expected. In line with our expectations we found that the increasing frustration in the form of the increasing value of $|R|$ results in decrease of the transition temperature. Further, we confirmed that at least for $R \geq -0.2$ the phase transition remains second-order and belongs to the standard Ising universality class. Nevertheless, our standard MC analysis became unreliable for $R < -0.2$ due to enormously increased equilibration and autocorrelation times near the transition point due to the freezing to domain states separated by large energy barriers and characterized by very slow dynamics. \\
\hspace*{5mm} The tunneling through the energy barriers was considerably improved by using the PT method, nevertheless, the convergence to the equilibrium still remained unreachable within accessible simulation times. Intuitively, it seems that here some cluster update methods~\cite{wolf89} might be useful, albeit, there is a report on their failure in the case of a frustrated $J_1-J_2$ system on the square lattice~\cite{kalz08}. Thus, the transition order in the highly frustrated region of $-1/4 <R < -0.2$ remains to be determined. Even though we did not see any convincing signatures of the first-order transition, one cannot overlook a relatively sudden drop in the internal energy as $R$ approaches $-1/4$ (see Fig.~\ref{fig:e-T}) even for a rather small lattice size. This might indicate a discontinuous behavior, and thus the first-order transition, in the thermodynamic limit but only a reliable FSS analysis (unachievable in the present study) could confirm or refute this speculation.\\
\hspace*{5mm} Another interesting open question is the existence and nature of the phase transitions in this model for $R < -1/4$. Unlike on the square lattice, the EFT found no LRO~\cite{boba16} but our recent MC study suggested a phase transition to a highly degenerate state consisting of frozen domains with the stripe-type AF ordering inside the domains separated by zero-energy domain walls~\cite{zuko20}. A phase transition for $R < -1/4$ was very recently reported also by using the machine learning technique~\cite{acev21}, however, the nature of the transition remains elusive.

\section*{Acknowledgments}
This work was supported by the grant of the Slovak Research and Development Agency under the contract No. APVV-16-0186 and the Scientific Grant Agency of Ministry of Education of Slovak Republic (Grant No. 1/0531/19).

\end{document}